\def\mn{ MNRAS}
\def\apj{ApJ}
\def\aps{ApJS}
\def\anr{ARA\&A}
\def\aaa{A\&A}
\def\aas{A\&AS}

\def\asj{AJ}

\def\nat{Nat}

\def\mn{MNRAS}
\def\et{et al.}

\def\msu{M$_{\odot}$}
\def\kms{km s$^{-1}$ }

\documentstyle[11pt,aaspp]{article}

\begin{document}

\title{Stripped Spiral Galaxies as Promising Targets for the Determination of the
Cepheid Distance to the Virgo Cluster}
\author{H. B\"ohringer, D.M. Neumann, S. Schindler\altaffilmark{1}}
\affil{Max-Planck-Institut f\"ur extraterrestrische Physik, D-85740 Garching,
Germany}

\author{J.P. Huchra}
\affil{Center for Astrophysics, Harvard University, 60 Garden Street,
MA 02138}

\altaffiltext{1}{also at Max-Planck-Institut f\"ur Astrophysik,
D-85740 Garching,Germany}

\begin{abstract}
The measurement of precise galaxy distances by Cepheid observations 
out to the distance of the Virgo cluster is important for the determination
of the Hubble constant. The Virgo cluster is thereby often used as an 
important stepping stone. The first
HST measurement of the distance of a Virgo galaxy (M100)
using Cepheid variables provided a value for $H_0 = 80 (\pm 17)$
\kms (Freedman \et\ 1994). This measurement was preceeded by a ground 
based study of the Virgo spiral NGC4571 (Pierce \et\ 1994) formally
providing $H_0 = 87 \pm 7$ km s$^{-1}$.
These determinations rely on the accuracy with which the
position of this observed spiral galaxy can be located
with respect to the Virgo cluster center.
This uncertainty introduces a major error in the
determination of the Hubble constant, together with the uncertainty 
in the adopted Virgo infall velocity of the Local Group.

Here we propose the use of spiral galaxies which show clear signs of
being stripped off their interstellar medium by the intracluster gas
of the Virgo cluster as targets for the Cepheid distance measurements.
We show that the stripping process and the knowledge of the intracluster
gas distribution from ROSAT X-ray observations allow us to locate these
galaxies with an at least three times higher precision 
with respect to M87 than in the
case of other spirals like M100. 
The X-ray observations further imply that M87 is well centered
within the densest part of the intracluster gas halo of the Virgo cluster.
This seems to imply that M87 is approximately at rest in the 
central core of the cluster. 

There remains the problem, however, that there is a velocity
difference between M87 and the average velocity of the
Virgo galaxies in the larger halo region around M87. This may be explained by
the disturbance introduced by the galaxy M86, which is falling into
the cluster with high velocity from behind. The X-ray and optical observations
imply that M86 is associated with a group of galaxies and a larger dark
matter halo with a mass of up to about 10\% of the mass of the M87
halo, which would account for the velocity offset of M87 with respect
to the velocity average of about 100 to 150 km s$^{-1}$.  

The combination of these informations 
could enable us to locate the two stripped spiral galaxies
quite precisely within the Virgo cluster as a whole
and could greatly improve the determination of the Virgo cluster distance.

\end{abstract}


\section{Introduction}
The Virgo cluster is an important stepping stone in the determination of
cosmological distances and of the expansion parameter (Hubble constant)
of the Universe (e.g. Tully 1988, Tammann 1991). 
An important objective of the Hubble Space
Telescope mission is the extension of the determination of cosmological
distances by means of the period-luminosity relation of Cepheid variables
out to the distance of the Virgo cluster of galaxies. Recently, the first
HST Cepheid distance measurement to the Virgo galaxy M100 was reported
(Freedman \et\ 1994, Mould \et\ 1995, Ferrarese \et\ 1996). 
The distance of the galaxy
was determined to be $17.1 (\pm 1.8)$ Mpc and a Hubble constant of
$H_0 = 80 (\pm 17)$ km s$^{-1}$ Mpc$^{-1}$ was deduced from this result.
The most recent update on this results by Ferrarese \et\ 1996 gives
a smaller distance of $16.1 \pm 1.3$ Mpc and a slightly higher
implied Hubble constant. Similarly in an earlier ground based study of the
Virgo spiral NGC4571 by Pierce \et\ (1994) a Cepheid distance of
$ 14.9 \pm 1.2$ Mpc was reported with a formally derived value of
$H_0 = 87 \pm 7$ \kms. 

How the distance of M100 is related to the distance of the Virgo cluster
as a whole is a very crucial step in this derivation of the Hubble constant.
Since the mean expected peculiar motion of M100
is too large to yield a precise measure of $H_0$ directly, the Cepheid
distance measurement is used to infer the distance of the Virgo cluster.
To obtain a value for $H_0$ the recession velocity of the Virgo cluster
(corrected for the ``Virgo infall'') is either used directly or the recession
velocity of the Coma cluster is combined with the relative distance
of the Virgo cluster and the Coma cluster in the paper by Freedman \et\
(1994) to derives values of of $H_0 = 82 (\pm 17)$ and 
$77 (\pm 16)$ km s$^{-1}$ Mpc$^{-1}$, respectively. An alternative
and very promissing way
to derive the ``cosmic recession velocity'' of the Virgo cluster for
this application is outlined by Jerjen \& Tammann (1993). They use 
absolute distance measurements to 17 clusters to establish a ``Machian
frame'' which allows to separate the Virgo recession and the Local
Group peculair motion. With the velocity values given by Jerjen \&
Tammann a lower value for $H_0$ is actually derived.  

Appart from this latter problem of separating the peculiar motion from the
cosmic expansion in the local Universe, the establishment of a proper
value for the Virgo distance is the most important, uncertain step
in the derivation of $H_0$ as described above. In the following we will 
therefore discuss if this determination can be improved. 
Using a Cepheid distance
measurement to obtain the Virgo distance one has to  
make the assumption that the observed spiral galaxy, e.g. M100, 
has a certain line-of-sight distance
from the center of the Virgo cluster. Freedman \et\ (1994) in particular assume
that M100 is located in the cloud of spiral galaxies centered on the Virgo
cluster with a line-of-sight extension of 6 Mpc. The lack of information on
the precise location of M100 thus adds to the uncertainties of the
determination of $H_0$ a value of 17.5\% in the paper by Freedman \et\ (1994).
This assumption is made somewhat {\it ad hoc}. Spiral galaxies are not
part of the central region of the Virgo cluster but rather avoid the
densest regions (Dressler 1984, Schindler \et\ 1995).
Mould \et\ (1995) for example show that a Virgo distance of even as far as
20 Mpc is still just consistent with the measured M100 Cepheid distance for
an approximately 90\% confidence interval. Therefore the inability to
locate precisely the galaxy with measured Cepheid distances within the Virgo
cluster is a severe limitation to the determination of the distance to
the Virgo cluster.

In the meantime also other Cepheid distances to Virgo cluster spirals
have been derived
using the Hubble Space Telescope. The galaxies NGC 4496, NGC 4536, 
and NGC 4639 were observed by Saha \et\ (1996a, 1996b) and  
Sandage \et\ (1996), respectively. The discovery of Cepheids in  NGC 4548 
was reported by Sakai \et\ (1996) with no distance values published yet. 
Taken together with the earlier observation of NGC4571 the galaxy distances 
stretch over a wide range: NGC4571 with $D = 14.9 \pm 1.2$ Mpc 
(Pierce \et\ 1994), M100 with $D = 16.1 \pm 1.3$ Mpc (Ferrarese \et\ 1996),  
NGC4496 with $D = 16.1 \pm 1$ Mpc (Saha \et\ 1996a), NGC 4639 with
$D = 17.8 \pm 0.4$ Mpc (Saha \et\ 1996b) and NGC4639 with $D = 25.1 \pm
2.8$ Mpc (Sandage \et\ 1996). While the good resolvability of the galaxy
NGC4571 (Sandage \& Bedke 1988) quite probably flags this object as
being in the foreground of the major part of the Virgo cluster, the other 
galaxies are believed to be true Virgo spirals as argued by the authors 
of the Cepheid observation papers (see also Tammann \et\ 1996). This 
highlights again the large extent of the galaxy infall region of
the Virgo cluster and the necessity of obtaining more precise
information on the location of such galaxies with respect to the structure
of the Virgo cluster.

In this paper we therefore like to discuss how galaxy targets for Cepheid
measurements can be selected, for which the relative position to the
center of Virgo are known more precisely. The selection of these targets is
based on the fact that interaction effects of the interstellar medium
with the intracluster gas of Virgo can be observed for some cluster spirals.
In particular the two galaxies NGC4548 and NGC4501 are observed to be
stripped off the interstellar gas (Cayatte \et\ 1990, 1994). In connection
with the knowledge of the gas density distribution in the Virgo cluster
from ROSAT X-ray observations and theoretical considerations on the
physics of the stripping process, the location of these two galaxies can be
confined to the surface of a
certain isodensity shell of the intracluster medium around M87.
The galaxy location can be well confined to a line-of-sight distance
from M87 to within $\pm 1$ Mpc. This is about a factor of three more
precise than in the case of M100.

The Virgo cluster as a whole is unfortunately quite complex, as observed in
the optical galaxy distribution (e.g. Binggeli \et\ 1987) and in X-rays
(B\"ohringer \et\ 1994), and it is therefore difficult to get precise
information on the line-of-sight location of M87 within the cluster.
In the optical M87 is located near the center of the densest, northern
part of the Virgo cluster, designated cluster A in Binggeli \et\ (1987).
The X-ray observation of the Virgo cluster in the ROSAT All Sky Survey
provides some further clues on this problem, implying that M87 is probably
centered in the halo of cluster A with some disturbance from the infall of the
M86 galaxy group, as will be discussed in this paper. 
This helps to relate the redshift of M87 to the recession velocity of
the Virgo cluster. In this way one can now relate the positions of
the stripped galaxies at the edge of the intracluster gas halo around M87
to the recession velocity of M87 and the Virgo cluster to derive the
Hubble parameter with largely reduced uncertainties as induced by the  
substructure in the Virgo cluster.
We therefore propose that the two galaxies NGC4548 and NGC4501 should
be chosen as targets for the search of Cepheid variables to get a more
reliable distance of the Virgo cluster.

In the following we discuss the stripping of Virgo
cluster spiral galaxies and the density distribution of the gaseous
X-ray halo of M87 in section 2 and derive the line-of-sight distances of
NGC4548 and NGC4501 with respect to M87. In section 3 we analyse the position
of M87 with respect to the intracluster medium of the Virgo cluster.
The results are discussed and summarized in section 4.
Throughout the paper we will assume a distance to the Virgo cluster of 20 Mpc
and for the easy conversion of our results to other distance values we will
give the scaling of the important quantities with distance by the parameter
$ d_{20} = D_{Virgo}/20 $ Mpc.

\eject

\section{The stripping of spirals and their location within the Virgo cluster}

As first proposed by Gunn \& Gott (1972) spiral galaxies are stripped
from their interstellar medium when they pass close enough to the
core of a galaxy cluster and this effect was partly made
responsible for the lack of late type galaxies in the densest
regions of galaxy clusters (e.g. Dressler 1984).
In the Virgo cluster we can clearly observe this effect for example
by a comparison of the neutral hydrogen content of spiral
galaxies measured in the 21 cm radio line by van Gorkum \& Kotanyi (1985)
and Cayatte \et\ (1990)
and the distribution of the intracluster gas as observed in X-rays
(B\"ohringer \et\ 1994) in the ROSAT All Sky Survey (Tr\"umper 1993,
Voges 1992). Fig. 1 shows an overlay of the 21 cm measurement in
23 spirals by Cayatte \et\ (1990) on top of the X-ray image of the
Virgo cluster from the ROSAT All Sky Survey. The X-ray image displays
the photon distribution detected in the energy range 0.4 to 2.4 keV
and the image is smoothed with a variable Gaussian filter.
As already noted by Cayatte \et\ (1994) the galaxies closer to the
center of the cluster  -- marked by the giant elliptical galaxy M87 --
are almost devoid of neutral hydrogen whereas the extent of the neutral
hydrogen gas disks increases with the distance from the cluster in the
sample of galaxies observed. At an intermediate location the galaxies
show signs of distortion in their gas disks which can be interpreted as
interaction effects of the gas disks with the hot, X-ray emitting
intracluster medium (ICM) of the Virgo cluster. This effect seems to be most
evident for the two galaxies NGC 4501 and NGC 4548.

Fig. 2 shows a close-up picture of this region again as an overlay of the
21 cm image of the two spiral galaxies NGC 4501 and NGC 4548 to the
X-ray surface brightness image of the ICM of the Virgo cluster
from the ROSAT All Sky Survey.
The neutral hydrogen distribution in the two galaxies as seen in 21 cm
is clearly asymmetrical. The interstellar gas appears to be compressed
at the side of the galaxies facing the Virgo cluster center and the gas is
combed backwards in the outer parts of the galaxies. This is just what would
be expected from the effect of ram pressure stripping. The Virgo cluster
center marked by the galaxy M87 is by far the deepest gravitational potential
in the region. It is therefore certain that the galaxies are attracted by it
and thus -- if they are on their first infall -- they are currently moving
towards the cluster center. Therefore the side with which they are facing
M87 (located in the direction of the lower right corner in Fig. 2) 
is the front-side with respect to their peculiar motion. Bernoulli
forces and Kelvin-Helmholtz instabilities 
acting at the sides of the gas disks are responsible for the
peanut shapes of the gas disks. For further details see
the images and data in Cayatte \et\ (1990, 1994).

For the X-ray image displayed in Fig. 2 four obvious point sources in the
region -- which are visible in Fig. 1 -- have been removed to show only
the angularly resolved structures which very probably correspond
to features in the gas distribution of the Virgo ICM.
The ICM appears actually quite patchy in this area. It is striking that
there are two surface brightness knots close to both of the being-stripped
spiral galaxies. In both cases the knots are offset from the galaxy in
the direction of the motion. They constitute surface density enhancements
of about 30 - 40\% above the background subtracted cluster emission. They
correspond to about 2$\sigma $ enhancements in this heavily smoothed image
(Gaussian used for smoothing is $\sigma = 500$ arcsec). They have extents
of up to 50 - 70 kpc with all the internal structure being smoothed-out.
Since the X-ray intensity is mostly a function of the emission measure
in the line of sight (and also because the intracluster gas is approximately in
pressure equilibrium on these scales) the intensity fluctuations have to
correspond to density enhancements of the gas. One can roughly estimate that
local density enhancements with a scale of $\sim 50$ kpc and an amplitude of
the order of a factor of two would produce the observed effect.

Given the low significance and the badly resolved structure of these
features we can only speculate. But the features could indeed be explained by
density enhancements in the ICM due to the sweeping-up
effect of the infalling galaxies. While the cold gas of the spirals is being
stripped, the ICM is piling up in front of the galaxies.
This would be a further confirmation of the fact that these two galaxies
are strongly interacting with the intracluster gas of the Virgo cluster.

Next we can discuss the ram pressure stripping effect 
quantitatively. As already discussed by Gunn \& Gott 
(1972, see also Sarazin 1986, Cayatte \et\ 1994) this effect is  
determined by the balance of the ram pressure, $P_{rp}$,
of the ICM pushing the interstellar medium out off the spiral
galaxy and the gravitational force with which the galaxy binds the
interstellar medium. The ram pressure depends on the density of the
ICM, $\rho_{ICM}$, and the velocity of the galaxy, $v_g$ :

\begin{equation}
P_{rp} = \rho_{_{ICM}}~ v_g^2
\end{equation}

The restoring force per unit area due to the gravity
of the disk of the spiral galaxy is given by (e.g. Gott \& Gunn 1972):

\begin{equation}
F_{grav} = 2\pi~ G~ \Sigma_{\star} \Sigma_{ISM}
\end{equation}

where $\Sigma_{\star} $ is the gravitational 
surface mass density and $\Sigma_{ISM} $ the 
surface density of the interstellar medium of the stripped spiral galaxy.

The 21 cm images of the spiral galaxies in the paper by Cayatte \et\ (1990)
are given out to the $10^{20}$ cm$^{-2}$ hydrogen column density contour. 
Therefore we calculate the stripping effect for this outer radius. For the
mass density of the galactic disk we use the mass average inside this
radius given by flat rotation curves and the assumption of Keplerian orbits.

\begin{equation}
\Sigma_{\star} = v_{\star}^2~R^{-1}~~(2\pi G)^{-1}
\end{equation}

With this simplification we get

\begin{equation}
F_{grav} = 3.03\cdot 10^{-12} {\rm dyn~ cm}^{-2} \left({N_H \over 
10^{20} cm^{-2}} \right)~ \left({v_{\star} \over 200 km s^{-1} }\right)^2~
\left({R \over 10 kpc}\right) ^{-1}~~d_{20}^{-1}
\end{equation}

and 

\begin{equation}
P_{rp} = 2.81 \cdot 10^{-12} {\rm dyn~ cm}^{-2} \left({n_e \over
10^{-4} cm^{-3}} \right)~ \left({v_{gal} \over 1200 km s^{-1} }\right)^2~  
\end{equation}

where $N_H$ is the column density of the neutral hydrogen ISM as measured
by the 21 cm emission and $v_{gal}$ is the infall velocity of the
spiral into the Virgo cluster.
The information on the rotation curves can also be taken from 21 cm 
observations by Guhathakurta \et (1988). For the infall velocity of
the galaxies into the Virgo cluster we take a value of 1200 km s$^{-1}$
which will be justified below. The gas density in the center of the Virgo
cluster can be obtained from the observed X-ray surface brightness of
the X-ray halo around M87 (B\"ohringer \et\ 1994; Nulsen \& B\"ohringer 1995)
with an approximate value as a function of radius of

\begin{equation}
n_{_{ICM}} \sim 0.8 \cdot 10^{-4}~ d_{20}^{-1/2}~{\rm cm}^{-3}~ 
\left({r\over 1~ {\rm MPC}~~ d_{20} }\right)^{-1.38}
\end{equation}

where $n_{_{ICM}}$ is the electron density in the ICM. The properties are
scaled by a Virgo distance parameter 
$d_{20} = D_{Virgo}/20$ Mpc. Taking the values for the rotational velocity
from the rotation curves published by Guhathakurta \et\ (1988) and the extents of
the hydrogen disks to the $10^{20}$ cm$^{-2}$ contour from
Cayatte \et\ (1990) as listed in Table 1 one finds that the ram pressure 
is balanced by the gravitational force of the galaxy for an M87 distance of
$0.62 {+0.74 \brack -0.34}~ d_{20}^{1.36} $ Mpc for NGC 4501 
and $ 1.01 {+1.20 \brack -0.55}~ d_{20}^{1.36} $ Mpc for NGC 4548.
The values we
find are close to the values for the projected distances of the galaxies
from M87. For the calculations of the uncertainties we have assumed 
that the considerations that went into the calculation of the pressure
balance in the ram pressure effect may be uncertain by 
up to a factor of three. 

Since we know that NGC 4501 is falling in from the front and 
NGC 4548 from behind into the Virgo cluster core, 
the distances to M87 are constraint to
$ D(M87) = (0.68 - 1.36)~ d_{20}$ Mpc for NGC 4501 and $ D(M87) = (0.8 
- 2.2) d_{20}$ Mpc for NGC 4548. Likewise the line-of-sight distance
difference between M87 and the stripped galaxies is $\Delta d(m87) =
(-1.18 - 0)~ d_{20}$ Mpc for NGC 4501 and $\Delta d(m87) = (0 - 2.07)~ d_{20}$
Mpc for NGC 4548. Therefore within this galaxy stripping scenario the
line-of-sight distance of the galaxies is fixed with respect to M87
with uncertainties of only one and two Mpc, respectively. This is a great
improvement of the positional uncertainty in view of the use
of these galaxies as probes for the Cepheid distance scale.

\begin{table*}
\begin{center}
\begin{tabular}{llll}
parameter & (units)&NGC 4501 & NGC 4548 \\
\tableline
radius &(arcmin)     & 2.5  & 2.4 \\
radius &(kpc)        & 14.3 $d_{20}$ & 13.7 $d_{20}$ \\
$v_{\star}$ &(km/s)   &  290  & 200 \\
$F_{grav}$ &(dyn/cm$^2$) & $4.4\cdot 10^{-12}~ d_{20}^{-1}$ &
  $2.2\cdot 10^{-12}~ d_{20}^{-1}$\\
$n_e(ICM)$ &(cm$^{-3}$) & $1.3\cdot 10^{-4}~ d_{20}^{-0.5}$ 
& $1.04 \cdot 10^{-4}~ d_{20}^{-0.5}$ \\
$D_{proj}$ &(Mpc) &  0.68 $d_{20}$ & 0.82 $d_{20}$ \\
R(M87)${^a)}$ &(Mpc)     &  0.62 $d_{20}^{1.36}$ & 1.01 $d_{20}^{1.36}$ \\
error of R${^b)}$ &(Mpc) & $ {+0.74 \brack -0.34}~ d_{20}^{1.36}$ &
$ {+1.20 \brack -0.55}~ d_{20}^{1.36}$ \\ 
\end{tabular}
\tablenotetext{a}{ Spatial distance to M87}
\tablenotetext{b}
{error in the distance calculation allowing for a factor of 3 uncertainty
in the description of the ram pressure effect}
\end{center}
\end{table*}

\begin{figure}[t]
\plotone{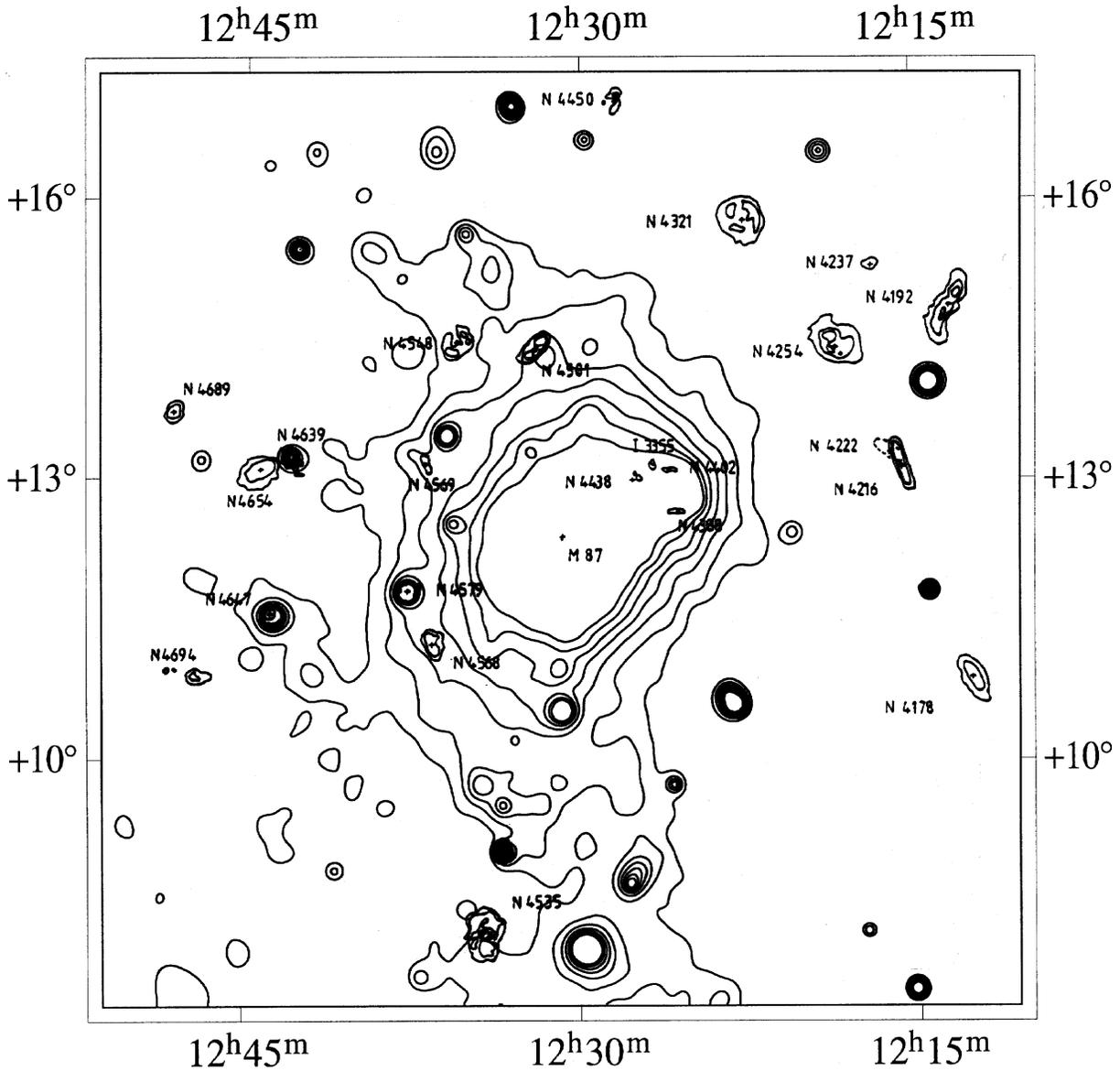}
\caption{Contour plot of the surface brightness distribution of the 
X-ray emission in the Virgo cluster as observed in the ROSAT All Sky Survey.
For the image only photons in the energy channels from 41 to 240 ($\sim 0.5 - 2$
keV) have been used and the image was smoothed with a variable Gaussian filter
(see also B\"ohringer \et\ 1994). The contour plots of the 21cm emission 
observed in 23 Virgo spirals in the survey by Cayatte \et\ (1990) are
superposed to the X-ray image. The X-ray contours are recognized by their
smoothness in this overlay. The contours in the inner part of the M87 halo
are not shown, since they would be so densly spaced to obscure all other marks.}
\end{figure}

\begin{figure}[t]
\plotone{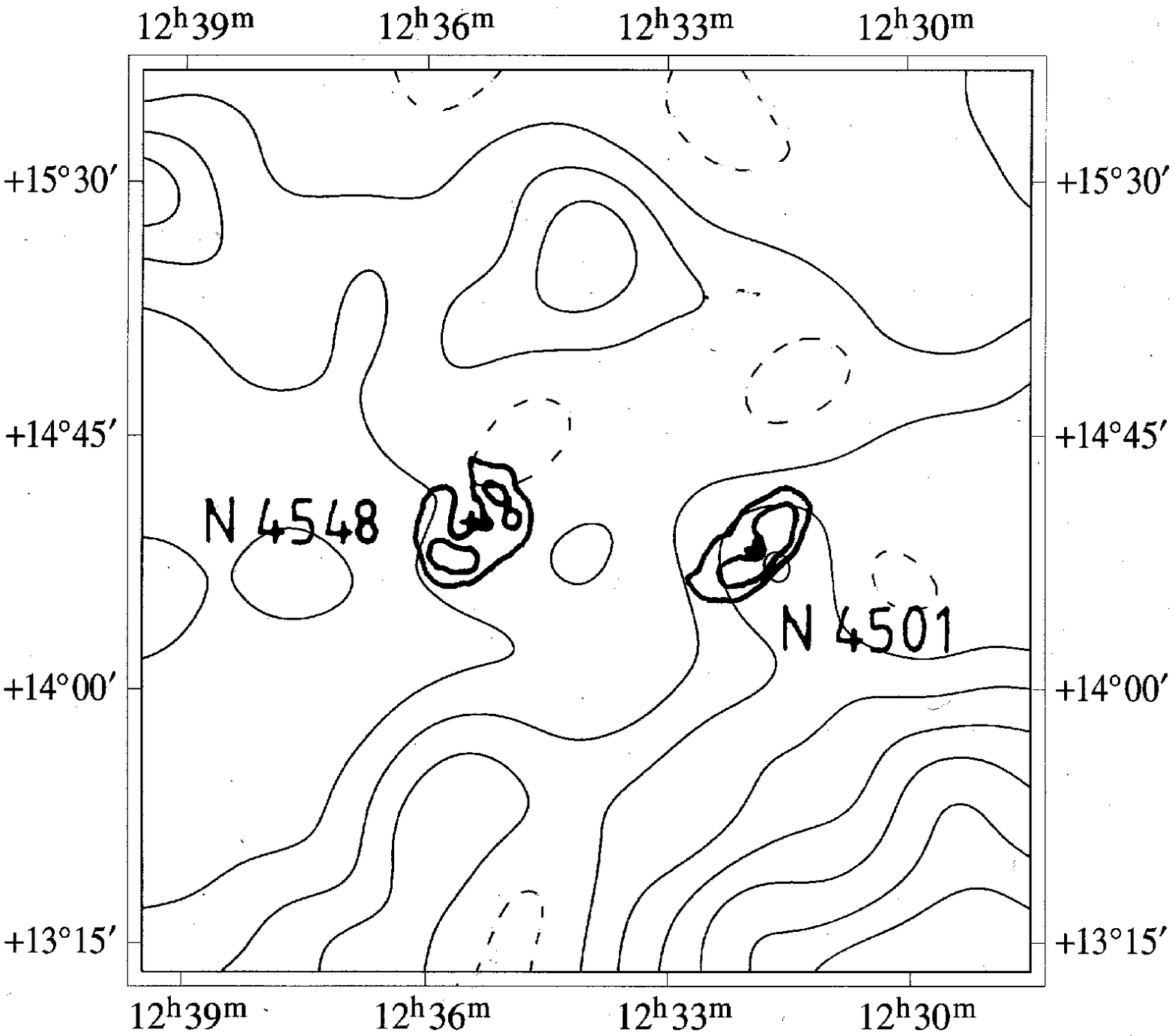}
\caption{Enlargement of Fig. 1 showing the surface brightness distribution
of the Virgo ICM in the region of NGC 4501 and NGC 4548. 
Clearly detected point sources have been removed from the data
before the photon distribution was heavily smoothed by a Gaussian
filter with $\sigma $ = 8.33 armin.
The 21 contour
maps of the galaxies are again superposed as heavy contour lines.
The X-ray emission is very patch in this region. But the two excess
emission regions in front of the two galaxies are clearly noticeable.
The dashed contours encircle minima in the X-ray surface brightness.}
\end{figure}

\section{M87 and its intracluster medium environment}

Another important consideration in fixing the Hubble constant with distance
measurements in the Virgo clusters is the question of how to determine
the recession velocity of the Virgo cluster. Fig. 1 shows clearly that the 
overall structure of the Virgo cluster is quite complex. In the X-ray image 
we observe this complexity in the plane of the sky. In redshift surveys
this is seen in the form of discrete sheets of galaxies with differing
redshifts. Thus there is a severe problem to define what is taken as
the Virgo cluster, or more specifically what should be considered as 
the relaxed part of the Virgo cluster. This problem and its relation
to the determination of $H_0$ has been discussed intensively in the
past (e.g. Tully 1988, Tammann 1991).

The ROSAT X-ray observations give some further help to the 
possible solution of this problem. In our analysis of the Virgo cluster
X-ray image from the ROSAT All Sky Survey (B\"ohringer et al. 1994) we
found that the main part of the surface brightness 
distribution can be quite well described
by a spherically symmetric halo component of hot plasma centered on M87.
This X-ray halo emits about 80\% of the total X-ray emission from the
Virgo cluster as shown in Fig. 1. M87 is well centered in this halo
which can be traced out to a radius of about 1.5 $d_{20}$ Mpc radius. 
A stronger deviation from
the spherically symmetric halo is only seen in the South-West where
the X-ray emission decreases very steeply. From this we concluded that
the Virgo cluster consists of an old, relaxed cluster core around M87
with a mass of $1.5 - 5.5 \cdot 10^{14}$ M$_{\odot}$ 
inside a radius of about 6 degrees and a larger
region of matter surrounding the core. X-ray halos around M49 and M86
are also clearly seen, but an order of magnitude less bright than the 
one of M87. The mass estimated for the relaxed core of the cluster
implies that it makes up an interesting fraction of the total mass of
Virgo estimated from the measured ``Virgo infall'' velocity (of
about 250 \kms ) to amount to about $10^{15}$ M$_{\odot}$. Therefore
a substantial mass fraction has to reside on this relaxed core of
the Virgo cluster.  

In a deep ROSAT PSPC pointing of about 30 ksec (for details see Nulsen
\& B\"ohringer 1995) we can analyse the relation of M87 with the X-ray
halo in more detail. A hard band image of the pointed observation of M87
is shown in Fig. 3. 
The first fact that one realizes is that both,
the X-ray halo and M87, show significant ellipticity. Carter and Dixon (1987)
have studied the elliptical shape of M87 in the optical and found a 
position angle of PA = 160$\pm 5$ degrees for the radial range 
of $0.5 - 26$ arcmin. The ellipticity is increasing from about 0.08 to 0.3
in the range from 0.5 to 5 armin and stays within 0.3 to 0.5 for larger
radii out to a radius of 25 arcmin. 

In Figs. 4 to 6 we show the results of a 
morphological analysis of the X-ray halo of M87.
The results were obtained by fitting ellipses to different isophote levels
of the X-ray image of M87 (for more details on the method see  
Bender \& M\"ollenhoff 1987; Neumann 1997). The position angle 
for the halo between 4 and 12 arcmin falls exactly in the same range
(158 $\pm 10$ degrees) as the one obtained for the optical image. 
In the inner region
for radii smaller than 4 arcmin the PA is less well defined. Also 
in this region (inside $\sim 5$ arcmin) we find complex 
features of substructure in
the surface brightness that are correlated with the radio lobes of
M87 as described in B\"ohringer et al. (1995) which we have cut out
from the image for the ellipticity analysis. This is part of the reason
why the PA becomes obscured in the center. The ellipticity over the same
radial zone ranges from 0.1 to 0.16. The ellipticity  
of the X-ray halo is thus smaller
than that for the light distribution in the central galaxy. This is to be
expected, because the gravitational potential is spanned by stars (and
probably by some form of dark matter) with anisotropic velocity dispersions
which create the elliptical potential shape. The hot, X-ray emitting gas 
of the halo is naturally isotropic in the random motions and thus traces
the potential in a more spherical way than the above matter components.
In addition the X-ray surface brightness is proportional to the emission 
measure of the hot gas (the line-of-sight integral of the gas density squared)
and therefore gives more weight to the center than to the regions at 
larger radii. This makes the X-ray halo more round. This effect has been 
discussed quantitatively by Buote \& Canizares (1992,1996) for a number
of elliptical Abell clusters. They find very similar values for the ratios of
X-ray and optical ellipticities as we find it here for M87 and they
show that these results can consistently be described by a model where
the optical luminous mass and the X-ray emitting gas are in equilibrium
within the same gravitational potential.
Therefore we take this as a sign that M87 is at rest within the gaseous
halo in the center of the Virgo cluster.
 
Finally, we can inspect the ellipse centers as a function
of the ellipse isophote level for the X-ray surface brightness
which is shown in Fig. 6. Here the isophote level has
been correlated to the corresponding major axis radius. One can see that
over a radius of 20 arcmin in East-West and North-South direction the
isophote ellipse centers shift by less than 1 arcmin. These are very 
small shifts compared to that observed in other clusters of galaxies 
(e.g. Mohr et al. 1993;
Allen et al. 1995). The center shifts
have been taken as a measure of substructure and the degree of non-relaxation
in the clusters. Thus compared to these other examples M87 is very well
centered in its X-ray halo which implies that M87 is almost at rest in
the center of the potential traced by the X-ray gas.

From these observations we would conclude that M87 is a very good reference
for the center of mass of the central, relaxed part of the Virgo cluster,
where it not for the fact that the radial velocity of M87 with a value
of 1258 \kms (heliocentric velocity) is quite different from the average
velocity of the galaxies in the cluster A component of Virgo with a value
of $1005 \pm 84$ \kms for E and S0 galaxies and $1123 \pm 75$ for 
dE and dS0 dwarf galaxies (Binggeli \et\ 1993). Thus we find a velocity 
offset of M87 with respect to the surrounding galaxies of
the order of 150 to 200 \kms. Inspecting again the X-ray image of Virgo
from the ROSAT All Sky Survey we note the bright X-ray halo around
the elliptical galaxy M86 which is close (1.28 degrees, 450 kpc in projection)
to M87. M86 has a blueshifted spectrum corresponding to a radial
velocity of -227 \kms (Binggeli \et\ 1993) and is thus falling into
the Virgo cluster from behind with a relative
velocity of at least about 1200 \kms. The impression of the distortion
of the hot gas halo of M86 due to the interaction with the Virgo core
that we get from the X-ray image of M86 (see e.g. Rangarajan \et\ 1995) 
leads to the conclusion that the interaction effects are near their
maximum and that therefore M87 and M86 are very close to each other and
have a similar distance.

M86 was also observed in a deep pointed observation with the
ROSAT PSPC (Rangarajan \et\ 1995) where the X-ray emission of the galaxy halo
can easily be traced out to 25 arcmin (150 kpc). A rough analysis of the
ROSAT All Sky Survey data (B\"ohringer \et\ 1994) and the deep ROSAT 
observation yield a mass of the X-ray emitting gas of about $ 2 - 3 \cdot
10^{11}$ \msu\ for a radius of 25 arcmin (where Rangarajan \et\ (1995) have
severely underestimated the gas mass of the M86 halo in their modeling).
Such a high gas mass is unusual for an isolated elliptical galaxy.
It corresponds much better to the gas masses found in the more massive of 
the compact Hickson groups (c.f. Ponman \& Bertram 1993; B\"ohringer 1994).
The interpretation of the X-ray halo of M86 as that of a group of galaxies
gets further support by the finding of Binggeli \et\ (1993) who note
that the distribution of the galaxy velocities in cluster A of Virgo
is quite asymmetric with a blueshifted tail. They note that the galaxies
in the tail which contain a large fraction of dE and dS0 dwarfs are 
associated with M86.  

Unfortunately, the X-ray halo of M86 is strongly distorted due to the 
merging of this galaxy group with the central part of the Virgo cluster and
it is therefore difficult to obtain a precise mass estimate of this group.
But if we tentatively reconstruct an undisturbed model halo for M86 by
extrapolating the X-ray parameters of the inner halo to a radius of
500 kpc (which is for example the outer radius to which the X-ray emission
could easily be traced in the more luminous Hickson groups HCG 62 and
HCG 97 (Ponman \& Bertram 1993; B\"ohringer 1994) and assume that the X-ray
gas is in approximate hydrostatic equilibrium for the average temperature
we observe in the halo today (about 0.9 keV) we find a mass of the order
of $2 \cdot 10^{13}$ \msu. This is surely the order of magnitude of the mass
we have to expect for this X-ray bright galaxy group around M86. 
We note that this amounts to about an order of magnitude less than the 
mass we find for the inner halo of M87. 

Because of the above mentioned interaction effects observed in the X-ray
image of M86 we can assume that the two galaxy systems associated with 
M87 and M86 are currently merging
and that we see the two systems in superposition/combination in cluster A.
Since it is almost impossible to distinguish the two systems and to assign
individual galaxies to either of the two, one can only obtain the center
of mass recession velocity of both by measuring the mean velocity of the
galaxies in cluster A. The velocity distribution for cluster A as shown
in Binggeli et al (1993) is clearly asymmetric and cannot be fit by a
single Gaussian, again underlining the merger configuration of cluster A.
In this context it is now easier to understand the velocity offset of
M87 with repect to the mean of all surrounding galaxies. If the M86 group
is attracted towards the M87 halo and has a mass of about 10\% of the
M87 halo mass and a relative velocity of 1200 km s$^{-1}$, a redshifted
velocity component of about 150 km s$^{-1}$ of M87 (and its associated 
galaxies) with respect to the common center of mass is just a consequence of
force balance. 
Thus if we would know the mass ratios of the two merging groups precisely 
we could account for the velocity
offset of M87 and could locate ther stripped spiral galaxies 
with respect to the Virgo cluster as a whole. This would be a promising
basis for determining a Cepheid distance for the Virgo cluster.

\begin{figure}[t]
\plotone{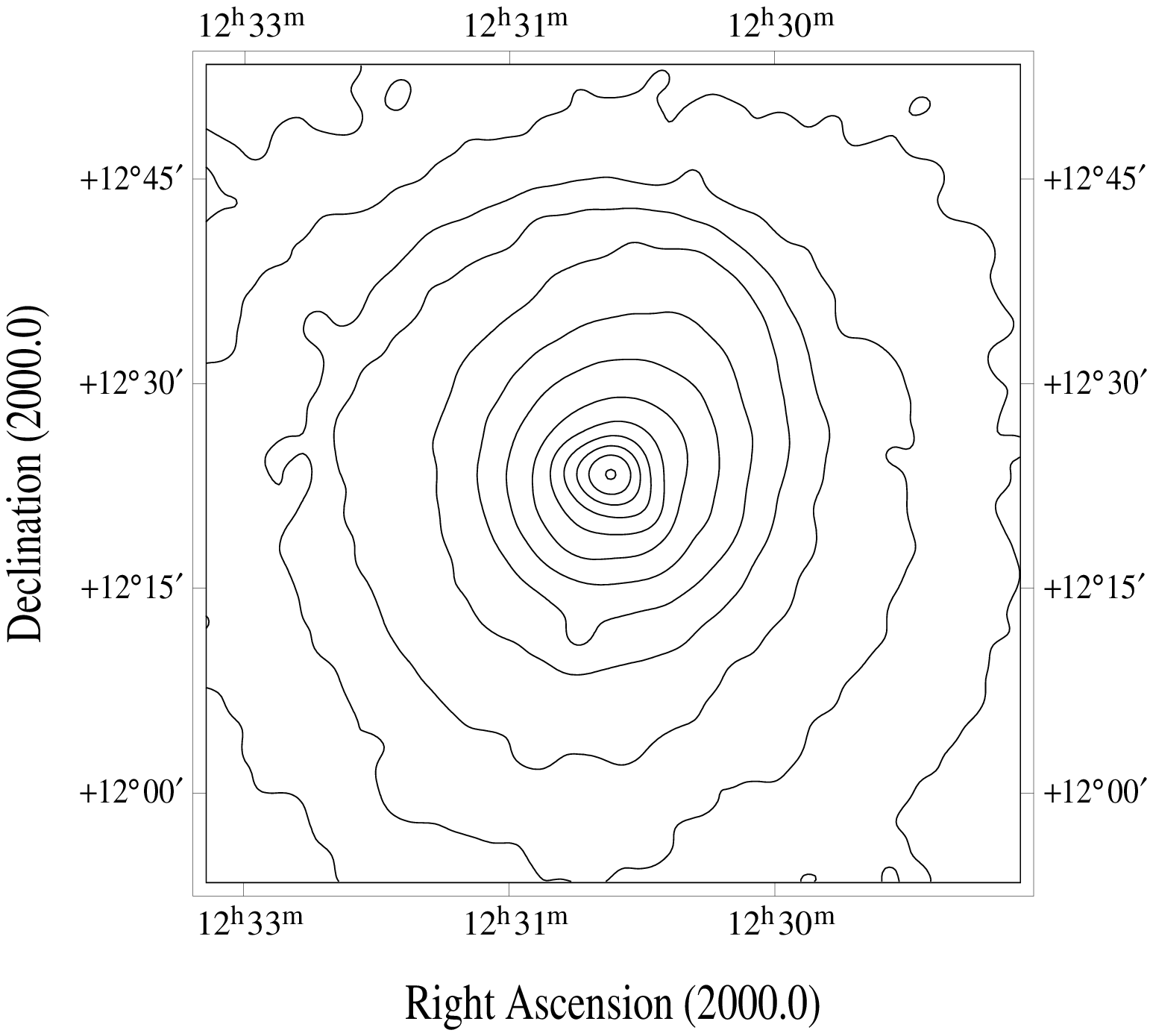}
\caption{Contour plot of the X-ray emission around M87 from the deep pointed
observation with the ROSAT PSPC. Only the photons in the energy chanels
from 52 to 201 are shown. The contours are clearly elliptical except for
the region inside a radius of 4 arcmin where the X-ray excess emission
of the radio lobes and the PSPC PSF obscure the ellipticity.}
\end{figure}

\begin{figure}[t]
\plotone{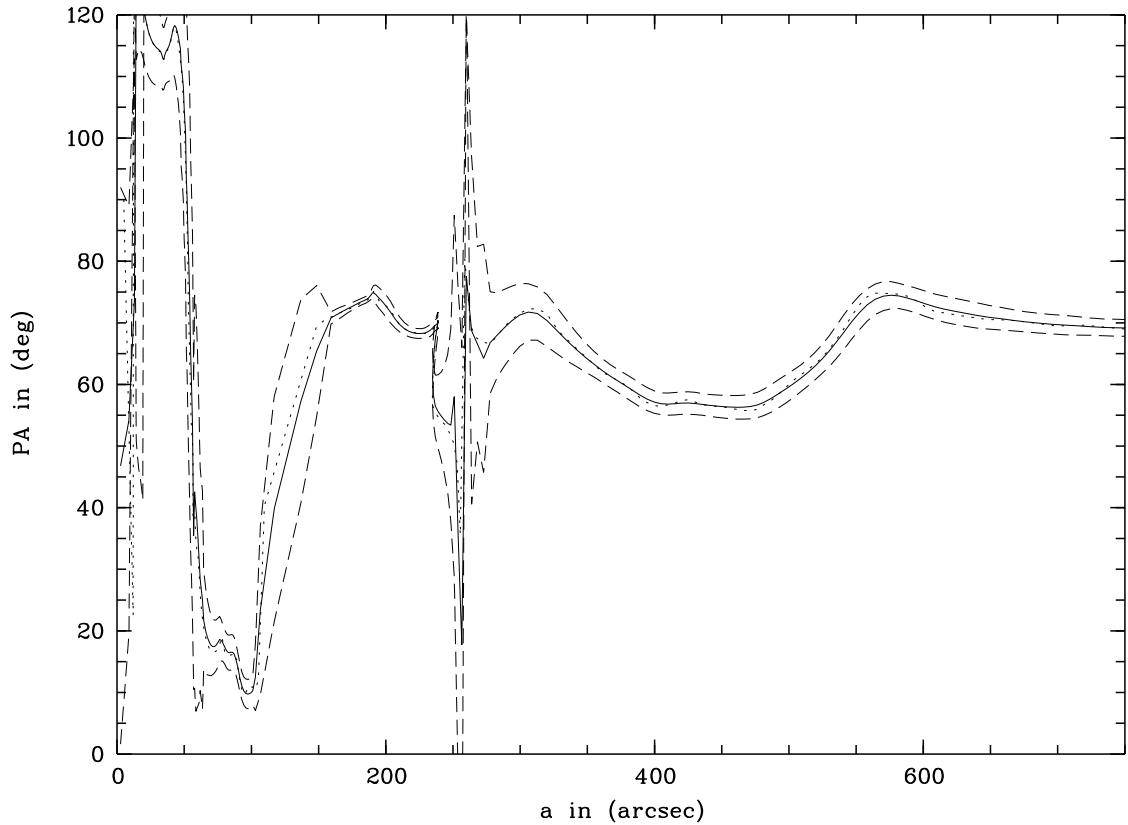}
\caption{ Position angle of the ellipticity of the X-ray surface brightness 
distribution in the halo of M87 as a function of the major axis radius.
The solid line shows the values derived from a fit to the observed data.
The dashed lines and the dotted line give the 10 and 90\% limits and the mean
for 100 Poisson realisations of the same observational data, respectively,
indicating the uncertainty of the parameter determination due to Poisson 
noise in the X-ray data.}
\end{figure}

\begin{figure}[t]
\plotone{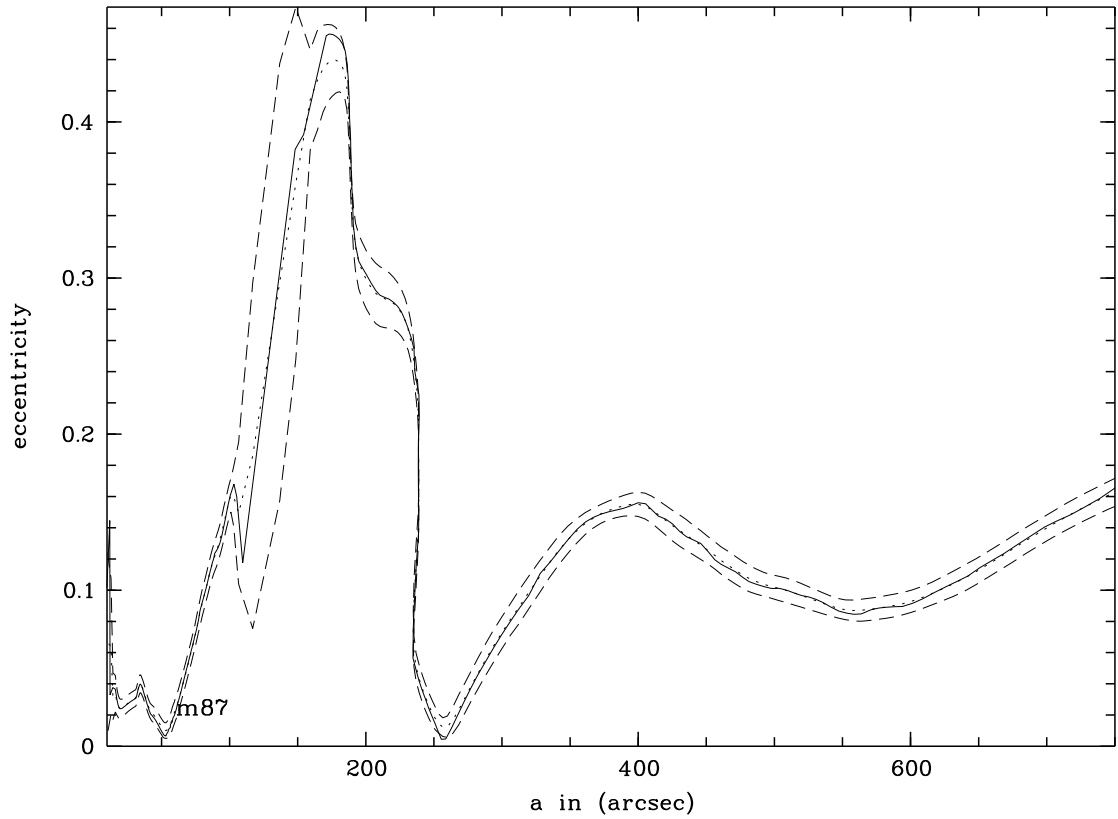}
\caption{Excentricity of the ellipticity of the X-ray surface brightness 
distribution in the halo of M87 as a function of the major axis radius.
The different lines have the same meaning as in Fig. 4}
\end{figure}

\begin{figure}[t]
\begin{tabular}{cr}
\plotone{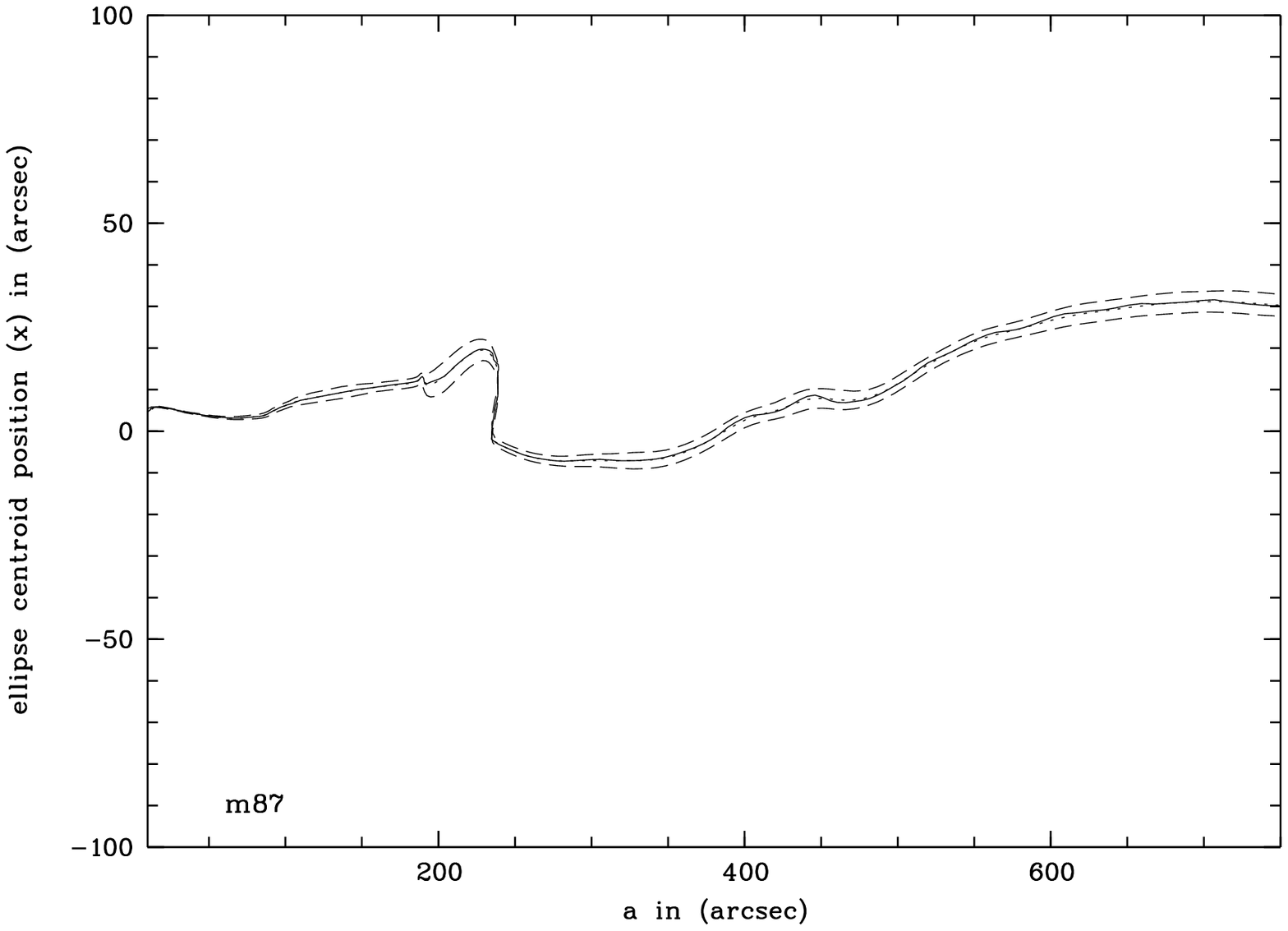} & \\
\plotone{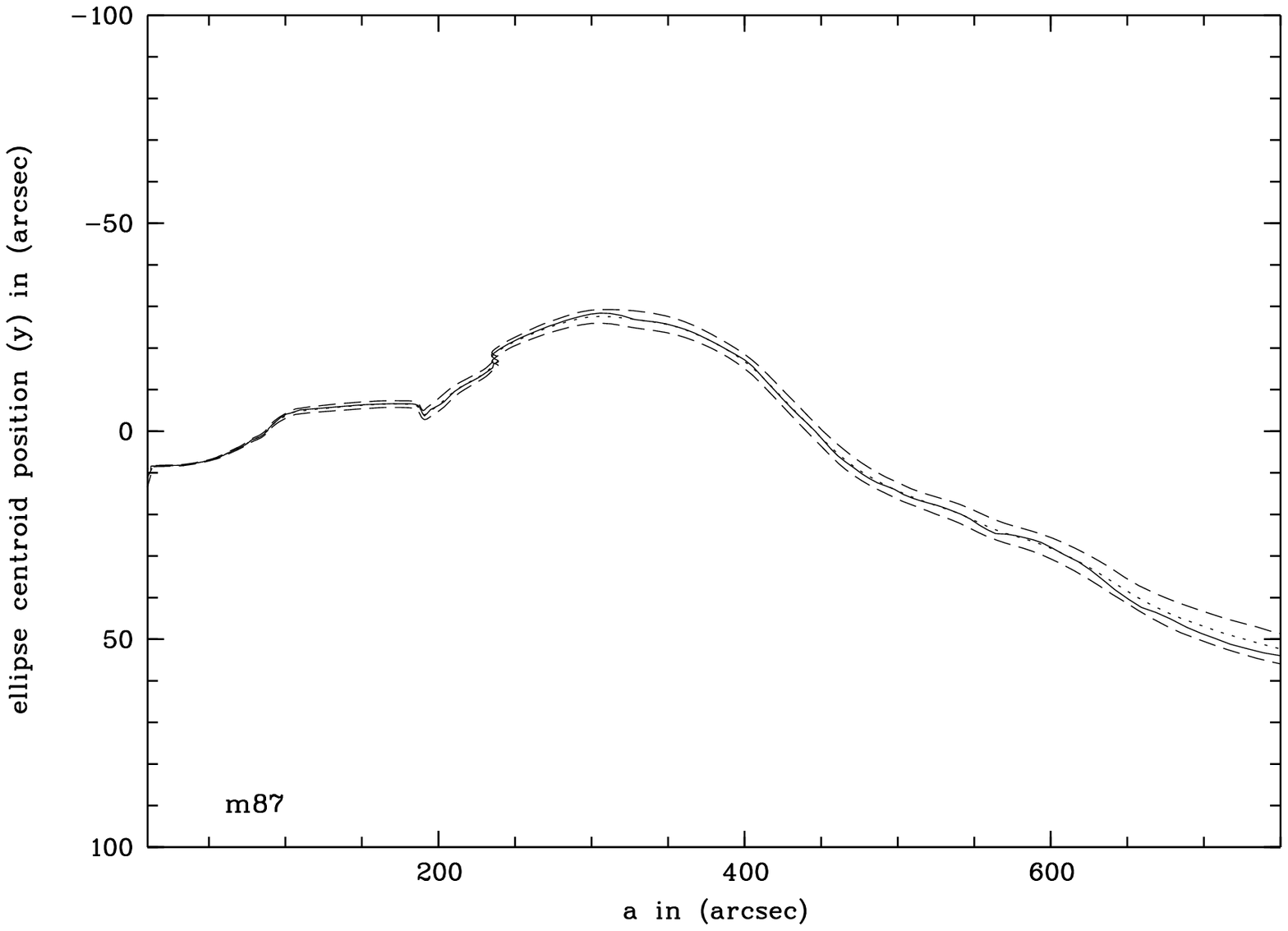} & \\
\end{tabular}
\caption{Centershifts for ellipses fitted to the isophotes of the 
X-ray surface brightness distribution in the halo of M87 as a function
of the major axis radius.
The different lines have the same meaning as in Fig. 4}
\end{figure}

\section{Discussion and Summary}

With the analysis of the above observations we have in principle 
established the 
following important link: We know the line-of-sight positions of two 
late type galaxies quite precisely with
respect to the location of M87 and we know that this 
galaxy is the center of the 
relaxed major mass component of Virgo. Now, to relate the recession velocity
of M87 to the velocity of the main body of the Virgo cluster we have to
correct for the influence of the infalling galaxy group M86. With the
positions of the galaxies NGC 4501 and NGC 4548 determined and a known
redshift of the central part of Virgo one has obtained a direct relation of
distance and recession velocity which would allow to determine the Hubble
constant (with the additional correction for the Virgo infall) 
in a more direct way than possible in the past. It is of course crucial
that both galaxies are of late type with a high enough star formation rate    
to make the search for Cepheids feasible. But since the stripping of the 
ICM is just starting now and the galaxies still have a dense interstellar
medium, there is no reason to expect a lower than normal population
of Cepheids in those galaxies. Both galaxies in particular NGC4501 
have impressive dust lanes visible in the optical images implying that
star formation should still be going on. 
NGC 4501 and NGC 4548 are in fact the 
galaxies closest in projection to M87 for which one can hope to find
Cepheids for a distance measurement. 
Therefore we propose that these two galaxies should be made targets of
Cepheid searches with HST with high priority (The observation of Cepeids 
in NGC4548 was actually just reported by Sakai \et\ 1996 while we were
working on a revision of this paper).   

In addition a better understanding of the dynamics within cluster A in
Virgo would be required. Of particular importance is a better determination
of the masses of the halo regions of M87 and M86. An improved and more
detailed analysis of the X-ray observations is in preparation by us. A
deeper analysis of the optical data is complicated by the fact that the
galaxy density in the immediate vicinity of M87 is surprisingly low.
If one would naively assume that the galaxy density is proportional to 
the density of the intracluster gas, one would expect a much larger number
of galaxies in the neighborhood of M87. Galaxy cannibalism has been
tentatively made responsible for this circumstance. This lack of galaxies is
so striking that for example the maximum of the projected galaxy density
is much closer to M86 than to M87 (see Binggeli \et\ 1987, Schindler
\et\ 1995). This uneven galaxy distribution makes it rather difficult
to use the galaxy redshift distribution in the A cluster to get a more
detailed picture of the M87-M86 merger.

We can further study the infall of the stripped galaxies in some more detail.
NGC 4501 has a recession velocity of 2283 \kms indicating that it is falling
in from the front, while NGC 4548 with a recession velocity of 483 \kms
is falling in from behind (Cayatte \et\ 1994, heliocentric velocities).
M87 has for comparison a recession velocity of 1258 \kms.
From the analysis of the X-ray data of the ROSAT All Sky Survey in 
the Virgo region (B\"ohringer et al. 1994) we get a rough estimate of the
mass profile of the M87 halo region with values in the range
$M(r) \sim 1.2 - 2.0\cdot 10^{14} 
\left( {r \over 1 {\rm Mpc}}\right) ^{1.0 - 1.4}$ \msu.
The arrival velocity of a galaxy in such a potential at a radius of
about 1 Mpc is in the range of 1200 to 1900 \kms. Here we have assumed that
the potential of the cluster has been constant over the relevant time, that is
at least for the last $\sim 5 $Gys, which may lead to a slight overestimate
of the galaxy velocity. The assumed value of the infall velocity in our
ram pressure stripping analysis is therefore in the right range. This
analysis can also be improved when better data on the mass distribution
in the M87 halo become available. This will be possible with better
information on the temperature distribution which will be obtained
from planned ASCA observations. 

There may also be other galaxies near the stripping radius that could be used
as targets after more detailed studies, for which the present data do not
show such clear evidence of stripping, like for example the galaxy
NGC 4569. For the modeling of the stripping effect it would be useful to have
even more detailed HI observations in those galaxies.

One should note that a good determination of the distance to Virgo
is not only important for a better knowledge of $H_0$, but is also 
interesting in itself. Virgo has been the target of so many studies
in particular in the field of environmental effects on the galaxy population
that a proper distance scaling of all the studied physical processes
would be a great advantage.

\acknowledgments

We thank Gustav Tammann, Bruno Binggeli and in particular Renzo Sancisi 
for fruitful discussions and helpful comments.
H.B. and S.S. are grateful
for financial support by the German BMFT through the Verbundforschung for
astronomy. S.S. gratefully acknowledges the hospitality of the Astronomical
Institute of the University of Basel.

\end{document}